\documentclass[apj]{emulateapj}

\newcommand{\hi}	  {\ion{H}{1}}
\newcommand{\hii}	  {\ion{H}{2}}
\newcommand{\kms}	  {km~s$^{-1}$}
\newcommand{\kmsm}	  {\mbox{ km~s}^{-1}}
\newcommand{\jone}        {$J=1\rightarrow0$}
\newcommand{\vlsr}	  {V_{\mbox{\tiny{LSR}}}}
\newcommand{\surfb}       {mag~arcsec$^{-2}$}

\slugcomment{Accepted for publication in \emph{The Astrophysical
Journal}}

\shorttitle{Cosmological Significance of HVC Complex H}
\shortauthors{Simon et al.}

\begin{document}

\title{The Cosmological Significance of High-Velocity Cloud Complex H}
\author{Joshua D. Simon\altaffilmark{1,2}, Leo Blitz\altaffilmark{1}, 
        Andrew A. Cole\altaffilmark{3}, Martin D. Weinberg\altaffilmark{4}, 
        and Martin Cohen\altaffilmark{5}}

\altaffiltext{1}{Department of Astronomy, 601 Campbell Hall,
                 University of California, Berkeley, CA 94720-3411;
                 jsimon@astro.berkeley.edu, blitz@astro.berkeley.edu}

\altaffiltext{2}{Current address: Department of Astronomy, MS 105-24,
                 California Institute of Technology, Pasadena, CA 91125;
                 jsimon@astro.caltech.edu}

\altaffiltext{3}{Kapteyn Astronomical Institute, University of Groningen,
                 Postbus 800, 9700 AV Groningen, the Netherlands;
                 cole@astro.umn.edu}

\altaffiltext{4}{Department of Physics and Astronomy, University of
                 Massachusetts, Amherst, MA 01003;
                 weinberg@astro.umass.edu}

\altaffiltext{5}{Radio Astronomy Laboratory, 601 Campbell Hall,
                 University of California, Berkeley, CA 94720-3411;
                 mcohen@astro.berkeley.edu}

\begin{abstract}

We have used new and archival infrared and radio observations to
search for a dwarf galaxy associated with the high-velocity cloud
(HVC) known as Complex~H.  Complex~H is a large ($\Omega \gtrsim
400$~deg$^{2}$) and probably nearby ($d = 27$~kpc) HVC whose location
in the Galactic plane has hampered previous investigations of its
stellar content.  The \hi\ mass of the cloud is $2.0 \times 10^{7}
(d/27\mbox{ kpc})^{2}$~M$_{\odot}$, making Complex~H one of the most
massive HVCs if its distance is more than $\sim20$~kpc.  Virtually all
similar \hi\ clouds in other galaxy groups are associated with low
surface brightness dwarf galaxies.  We selected mid-infrared sources
observed by the MSX satellite in the direction of Complex~H that
appeared likely to be star-forming regions and observed them at the
wavelength of the CO \jone\ rotational transition in order to
determine their velocities.  59 of the 60 observed sources show
emission at Milky Way velocities, and we detected no emission at
velocities consistent with that of Complex~H.  We use these
observations to set an upper limit on the ongoing star formation rate
in the HVC of $\lesssim5 \times 10^{-4}$~M$_{\odot}$~yr$^{-1}$.  We
also searched the 2MASS database for evidence of any dwarf-galaxy-like
stellar population in the direction of the HVC and found no trace of a
distant red giant population, with an upper limit on the stellar mass
of $\sim10^{6}$~M$_{\odot}$.  Given the lack of evidence for either
current star formation or an evolved population, we conclude that
Complex~H cannot be a dwarf galaxy with properties similar to those of
known dwarfs.  Complex~H is therefore one of the most massive known
\hi\ clouds that does not contain any stars.  If Complex~H is
self-gravitating, then this object is one of the few known dark galaxy
candidates.  These findings may offer observational support for the
idea that the Cold Dark Matter substructure problem is related to the
difficulty of forming stars in low-mass dark matter halos;
alternatively, Complex~H could be an example of a cold accretion flow
onto the Milky Way.

\end{abstract}

\keywords{Galaxy: evolution --- galaxies: dwarf --- Local Group ---
infrared: ISM --- infrared: stars --- radio lines: ISM}

\section{Introduction}
\label{chap3-intro}

\begin{figure*}[th!]
\plotone{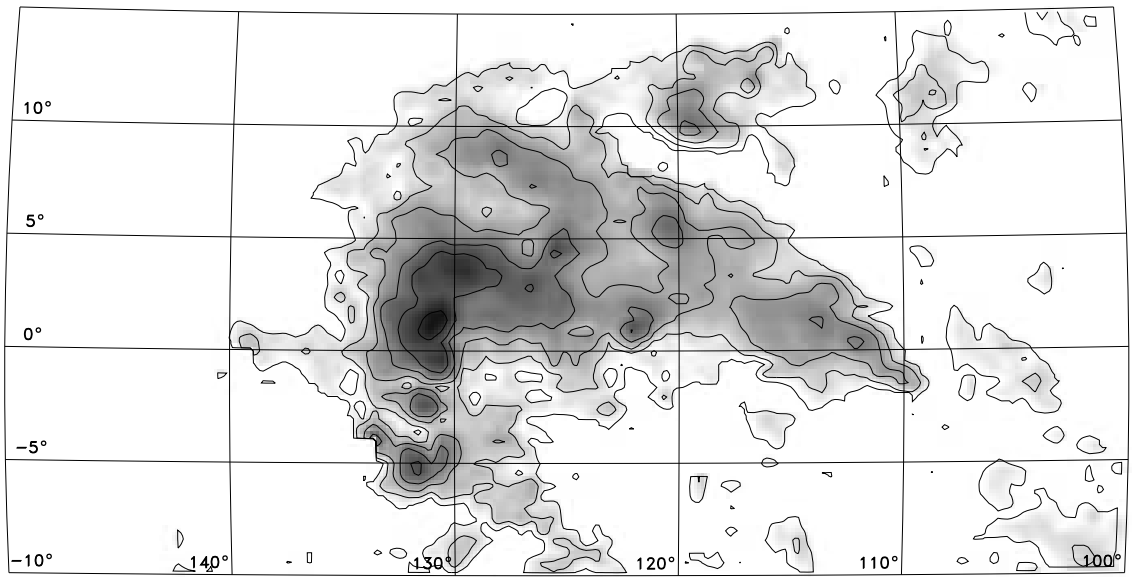}
\label{fig:chap3-ldsmap}
\caption{\hi\ map of Complex~H in Galactic coordinates from the
  Leiden/Dwingeloo Survey of Galactic Neutral Hydrogen \citep{dap}.
  The image shows the integrated intensity for velocities between
  $-230$ \kms\ and $-150$ \kms.  The contours are spaced
  logarithmically and correspond to column densities of $4.6 \times
  10^{18} \mbox{ cm}^{-2}$, $9.2 \times 10^{18} \mbox{ cm}^{-2}$, $1.8
  \times 10^{19} \mbox{ cm}^{-2}$, $3.7 \times 10^{19} \mbox{
  cm}^{-2}$, $7.3 \times 10^{19} \mbox{ cm}^{-2}$, and $1.5 \times
  10^{20} \mbox{ cm}^{-2}$.  The Dwingeloo telescope has a beam size of
  36\arcmin\ and the survey was carried out with a grid spacing of
  30\arcmin.}
\end{figure*}

What determines whether a low-mass gas cloud or dark matter halo
becomes a dwarf galaxy or fails to undergo any star formation,
remaining dark for billions of years?  The answer to this question may
underlie the substructure problem in Cold Dark Matter (CDM)
cosmologies --- the dramatic mismatch between the number of dark
matter minihalos produced in numerical simulations and the number of
dwarf galaxies observed in the Local Group \citep{klypin99,moore99}.
Many possible explanations for why low-mass halos might not form stars
have been proposed
\citep*[e.g.,][]{e92,evan00,bullock00,evan01,somerville,voj02,dw03,kgk04},
but the existence of very low-mass dwarfs today suggests that this
problem is not yet fully understood.  One approach to improving our
understanding of the formation of dwarf galaxies is to study extreme
objects in order to determine what makes them unique.  In this paper
we investigate the nature of Complex~H, an unusually massive
high-velocity cloud (HVC) located in the Galactic plane (see Figure
\ref{fig:chap3-ldsmap}).

Complex~H is centered on HVC~131$+$1$-$200, which was first discovered
by \citet{hulsbosch71} and \citet{dieter71}.  \citet{wvw} noted that
this HVC seems to be associated in position and velocity with a large
number of other clouds, and named the grouping Complex~H after
Hulsbosch.  The complex subtends 478 deg$^{2}$ \citep{wvw} and extends
in velocity all the way from $\vlsr=-230$ \kms\ down to $\vlsr \approx
-120$ \kms\ where it begins to blend into Milky Way emission.  The
\hi\ column density at the center of the cloud is $\gtrsim2\times$
$10^{20}$ cm$^{-2}$, and the column is at least 2.0 $\times$ $10^{19}$
cm$^{-2}$ for most of the central 25 deg$^{2}$.  Complex~H has an
integrated \hi\ mass of $\approx$ 2.7 $\times$ $10^{4}$
$d_{\mbox{\small{kpc}}}^{2}$ M$_{\odot}$, where
$d_{\mbox{\small{kpc}}}$ is the distance to Complex~H in kiloparsecs
\citep[hereafter W98]{w98}.  \citet{ic97} used IRAS data to search for
star formation in Complex~H and found one candidate young stellar
object, but this source is likely to be a foreground Milky Way object.

W98 examined ultraviolet spectra of 17 OB stars for interstellar
absorption lines of \ion{Mg}{2}, \ion{C}{2}, and \ion{O}{1} in the
direction of Complex~H.  They detected no absorption near the velocity
of Complex~H and concluded that even for substantially subsolar
abundances, the HVC must be located beyond the farthest of the stars
they studied.  They placed a firm lower limit of 3.4 kpc on the
distance to Complex~H.  Since the distances of some of the OB stars
are rather uncertain, the actual minimum distance could be as large as
6.5 kpc.  \citet{b99} additionally pointed out that the velocity of
Complex~H is too large for it to be in circular rotation around the
Galaxy at any distance.  They further argued that the lack of any
observational evidence for an interaction between Complex~H and the
interstellar medium of the Milky Way strongly implies that the HVC
must be located beyond the edge of the disk of the Galaxy, at least
$\sim20$ kpc away from the Sun.  Since Complex~H has velocities of up
to 100 \kms\ with respect to the nearest Galactic gas, strong shocks
and X-ray and radio emission would be produced if it were located
within the \hi\ disk of the Milky Way.  The most recent \hi\
observations of Complex~H by \citet{lockman} do indicate that the HVC
is beginning to interact with the Milky Way, but this interaction
takes the form of tidal stripping of Complex H's outer layers rather
than the high-energy collision that would be occurring if Complex~H
were closer than the edge of the disk.

\citet{lockman} further noticed that the velocity gradient across the
cloud in the $b$ (Galactic latitude) direction can be used to derive
its vertical motion relative to the Galaxy.  Using \hi\ maps from the
Green Bank Telescope, he constructed a model of Complex~H and argued
that the HVC is in an inclined, retrograde orbit around the Milky Way.
These calculations place Complex~H $33\pm9$ kpc from the Galactic
center and $27\pm9$ kpc from the Sun.  More general distance
constraints can be derived by considering the aforementioned lack of a
strong interaction between Complex~H and the Milky Way, and the total
mass of the cloud.  The Milky Way disk gas extends out to a minimum
Galactocentric distance of 27 kpc at the position of Complex~H
\citep{b99}, placing a firm lower limit of 21 kpc on the distance
between the HVC and the Sun if the cloud lies just beyond the edge of
the disk.  At distances of more than 100 kpc, Complex~H would be the
fourth most massive object in the Local Group, which seems unlikely
since no counterpart at other wavelengths has been detected.  The full
range of plausible distances is therefore $21-100$ kpc, corresponding
to \hi\ masses of $1.2 \times$ $10^{7} - 2.7 \times
10^{8}$~M$_{\odot}$.

Because starless extragalactic gas clouds as large as Complex~H are
not seen in other groups of galaxies (with only one exception
[\citealt{minchin05}]), it seems reasonable to suppose that Complex~H
is the \hi\ component of a previously undiscovered dwarf galaxy.  At
the distance of 27 kpc preferred by \citet{lockman}, Complex~H would
have an \hi\ mass, total mass (assuming that it is gravitationally
bound), and physical extent that are consistent with those of other
Local Group galaxies.

Before we proceed, the issue of the total mass of Complex H deserves
some comment.  If the HVC is actually a dwarf galaxy, a baryon
fraction of 0.04 (typical for dwarf galaxies) would imply a total mass
of $\sim5 \times 10^{8}$~M$_{\odot}$.  In a $\Lambda$CDM cosmology,
such a dark matter halo should have a peak circular velocity of at
least $\sim30$~\kms.  Neither the velocity gradient across the cloud
(which in the Lockman model is attributable to the orbital velocity of
the HVC rather than its internal motions) nor the velocity dispersion
of the gas is nearly this large, so the kinematics of the cloud do not
require a total mass this high.  Nevertheless, if the \hi\ extent of
the cloud is smaller than the scale radius of the dark matter halo,
the observed kinematics would not be expected to reflect the full
gravitational potential of the halo.

We now consider the possibilities for confirming or refuting the
presence of a dwarf galaxy in Complex~H.  The $V$-band extinction in
the direction of the center of the HVC is estimated to be 4 magnitudes
\citep*{sfd98}, although the calculated extinction near the Galactic
plane is subject to significant uncertainties.  The combination of
heavy extinction and severe crowding makes an optical detection of a
distant group of stars very difficult (although not impossible, as was
illustrated by the detection of the Sagittarius dwarf spheroidal
behind the Galactic center by \citealt*{ibata95}).  Instead, we shall
search for evidence of a dwarf galaxy at longer wavelengths where both
the extinction and the crowding are less severe or nonexistent.
Specifically, we use mid-infrared observations from the Midcourse
Space Experiment (MSX) satellite to identify star-forming regions that
could be associated with Complex~H.  We then employ millimeter-wave CO
observations to determine the nature and location of these objects.
These data should reveal the presence of massive stars in Complex~H if
any star formation has taken place in the last $\sim10^{7}$ years.
Other Local Group dwarf irregulars that have comparable \hi\ masses to
Complex~H, such as Sextans~B and IC~1613, are actively forming stars.
In addition, we use the Two Micron All Sky Survey (2MASS)
near-infrared database to search for stars in the putative dwarf
galaxy, taking advantage of the order of magnitude decrease in
extinction between $V$-band and $K$-band.

In the following section, we discuss the MSX dataset, present our new
CO observations, and describe the results of our search for star
formation.  In \S \ref{chap3-2mass} we describe our analysis of the
2MASS data and in \S \ref{chap3-discussion} we consider the
interpretation of our findings.  Our conclusions are summarized in \S
\ref{chap3-conclusion}.

\section{SEARCH FOR RECENT STAR FORMATION IN COMPLEX~H}
\label{chap3-msx}

\subsection{MSX Data}
\label{chap3-data}

The MSX satellite \citep{msx} was a US Department of Defense mission,
undertaken by the Ballistic Missile Defense Organization.  Launched in
1996, the infrared instrument on board consisted of a 35-cm off-axis
telescope with detectors in six mid-infrared bands (two very narrow
near 4 $\mu$m, and four broad at roughly 8, 12, 15, and 21 $\mu$m).
All sensors had pixels of 18.3\arcsec.  The primary infrared product
was a survey of the entire Galactic plane within $-5\degr < b <
5\degr$ and about 20\arcsec\ (FWHM) resolution \citep{price01}.  The
MSX Point Source Catalog version 2.3 (PSC2.3) contains six-color
infrared photometry for over $4.3 \times 10^{5}$ sources in the
Galactic plane \citep{msxps03}.  The MSX data products that we
use for our analysis have a 3~$\sigma$ sensitivity limit for low
surface brightness diffuse emission that varies across the field from
$2 - 9 \times 10^{-7}$~W~m$^{-2}$~sr$^{-1}$ and are sensitive to point
sources with flux densities down to $\approx100$~mJy.\footnote{We only
use the 8~$\mu$m data in this paper because of the much lower
sensitivity in the longer wavelength bands.}

\subsection{Detectability of Complex~H in the Mid-IR
\label{chap3-detectability}}

If Complex~H is indeed the gaseous counterpart of a dwarf galaxy, the
galaxy could in principle be either a dwarf irregular (dIrr) or
a dwarf spheroidal (dSph), although its gas content would be
rather large for a dSph \citep{br00}.  The photospheres of
individual stars in such a galaxy are much too faint to have been
detected with MSX, since even an O star would have a flux density of
less than 1~mJy at 8~$\mu$m for a distance of 27~kpc.  However, \hii\
regions containing warm dust can become much brighter; if dust in the
vicinity of an O star is heated to a temperature of more than
$\sim100$~K, it will be easily visible with MSX.  Emission from the
polycyclic aromatic hydrocarbon (PAH) bands at 6.2~$\mu$m, 7.7~$\mu$m,
and 8.7~$\mu$m in photodissociation regions could further enhance the
detectability of such star forming regions, although the likely
abundance of PAHs in this environment is quite poorly known.

Since all known dwarf irregular galaxies contain \hii\ regions (with
the exception of some of the transitional dwarfs), it is a reasonable
expectation that Complex~H might contain \hii\ regions as well.  In
order to estimate the brightness of these star-forming regions, we
used other Local Group dwarfs as a comparison.  NGC~6822, IC~10, and
the SMC have all been observed in the mid-infrared by MSX, the
\emph{Spitzer Space Telescope}, or both.  These galaxies have
approximately an order of magnitude more neutral gas than Complex~H,
but none of the lower-mass dwarfs has been observed yet at useful
sensitivities and resolutions.  Their metallicities are low, $Z = 0.1
- 0.3 Z_{\odot}$ \citep*{cbf00,l79,ptp76}, similar to what is seen in
HVCs, so these galaxies are plausible proxies for Complex~H.  We
examined five \hii\ regions in these galaxies: Hubble V and Hubble X
in NGC~6822, HL45 and HL106b \citep{hl90} in IC~10, and N81 in the
SMC.  When scaled to a distance of 27 kpc, Hubble V and Hubble X are
extremely bright, with 8~$\mu$m flux densities\footnote{The 8~$\mu$m
bands of \emph{Spitzer} and MSX have somewhat different shapes and
bandwidths, but their fluxes tend to agree within 20\%, so we do not
attempt any correction to put all of these observations on an
identical scale.} of 32 Jy and 10 Jy, respectively.  These two regions
do have relatively high star formation rates
($\sim10^{-3}$~M$_{\odot}$~yr$^{-1}$, estimated using the
\citeauthor*{kenn94} [1994] relation between H$\alpha$ luminosity and
star formation rate), making them less than ideal comparison objects.
The IC~10 \hii\ regions and N81 may be more appropriate templates,
with scaled 8~$\mu$m flux densities of 1.82~Jy, 3.25~Jy, and 0.81~Jy,
respectively.  The corresponding star formation rates are $7 \times
10^{-5}$~M$_{\odot}$~yr$^{-1}$ and $1 \times
10^{-5}$~M$_{\odot}$~yr$^{-1}$ in IC~10 (again calculated from the
\citeauthor{kenn94} relation) and $5 \times
10^{-4}$~M$_{\odot}$~yr$^{-1}$ in N81 (derived from \emph{Hubble Space
Telescope} observations of the OB stars in the region [\citealp{hm99}]
with extrapolation to low-mass stars from a \citeauthor{salpeter}
[1955] or \citeauthor{kroupa} [2001] initial mass function).  A final
point of comparison comes from MSX observations of some of the most
distant known molecular clouds in the outer Galaxy.  Of the 11 clouds
listed by \citet*{ddt94} at distances from the Sun between 10 and 21
kpc, 5 are detected by MSX, with 8~$\mu$m flux densities ranging from
108~mJy to 470~mJy.  This demonstrates that faint \hii\ regions at
similar distances to what we assume for Complex~H are detectable in
MSX data.

\subsection{Target Selection}
\label{chap3-targetselection}

The $20\arcsec$ resolution of the highest-resolution MSX data
corresponds to a physical size of $0.097 d_{\mbox{\small{kpc}}}$~pc,
where $d_{\mbox{\small{kpc}}}$ is the distance to Complex~H in
kiloparsecs.  A typical \hii\ region of diameter 50 pc would subtend
an angle of $6\farcm4$ at the nominal distance of 27 kpc to Complex~H,
and would thus be resolved easily by MSX.  Ultracompact \hii\ regions
(less than 2 pc in diameter), however, would show up as point sources
at distances of greater than $\sim20$~kpc.  Therefore, we consider
both point sources and extended sources in our search for evidence of
star formation in the HVC.

\begin{figure*}[th!]
\plotone{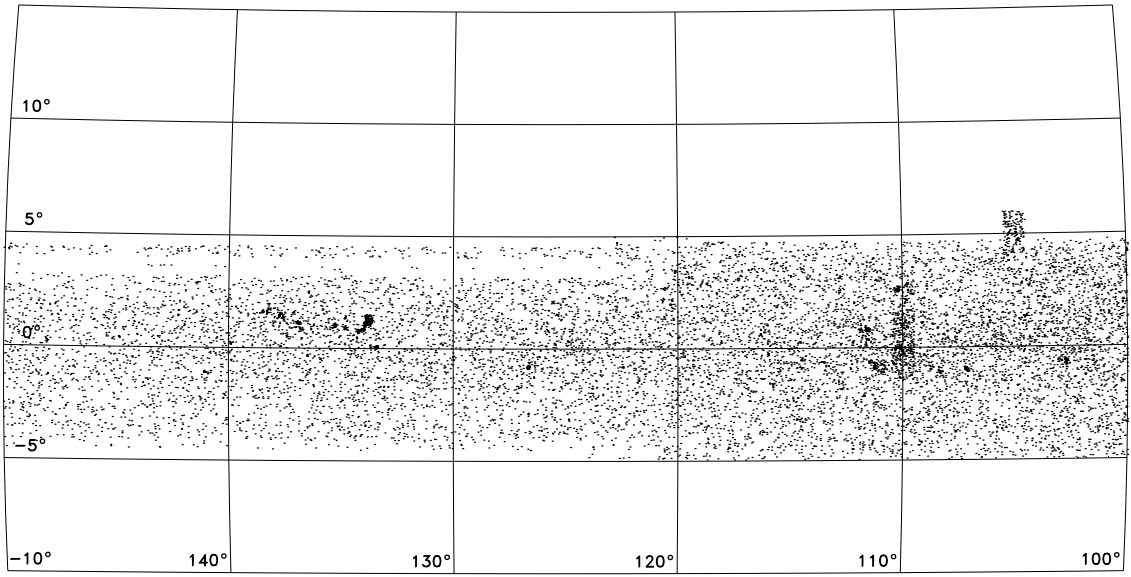}
\caption{MSX point sources in the outer Galaxy.  All sources listed in
the Point Source Catalog \citep{msxps03} are plotted.  Overdensities
near the position of Complex~H ($\ell = 131\degr, b = 1\degr$; see
Figure \ref{fig:chap3-ldsmap}) are visible, but they can all be
identified with nearby Milky Way \hii\ regions. }
\label{fig:chap3-msxpsc}
\end{figure*}

As a first step, we searched the PSC2.3 for evidence of an enhanced
density of infrared point sources near the center of Complex~H.  The
distribution of these sources in the plane of the Galaxy is displayed
in Figure \ref{fig:chap3-msxpsc}.  Although there are a few
overdensities close to the expected location, further investigation
reveals that the string of sources from $\ell \approx 132\degr -
138\degr$ is associated with the Milky Way star forming complex
W3/W4/W5 \citep*[e.g.,][]{chs00} and the cluster at $\ell$ =
126.7\degr, $b$ = $-0.8$\degr\ is the \hii\ region Sharpless~187
\citep*{sharpless59,joncas92}.  It is also worth noting that the
average density on the sky of 8~$\mu$m point sources around $\ell$ =
131\degr, $b$ = 1\degr\ (the core of the HVC) is very close to the
mean value for the outer Galactic plane of about 20 deg$^{-2}$.

\begin{figure*}[th!]
\epsscale{1.17}
\plotone{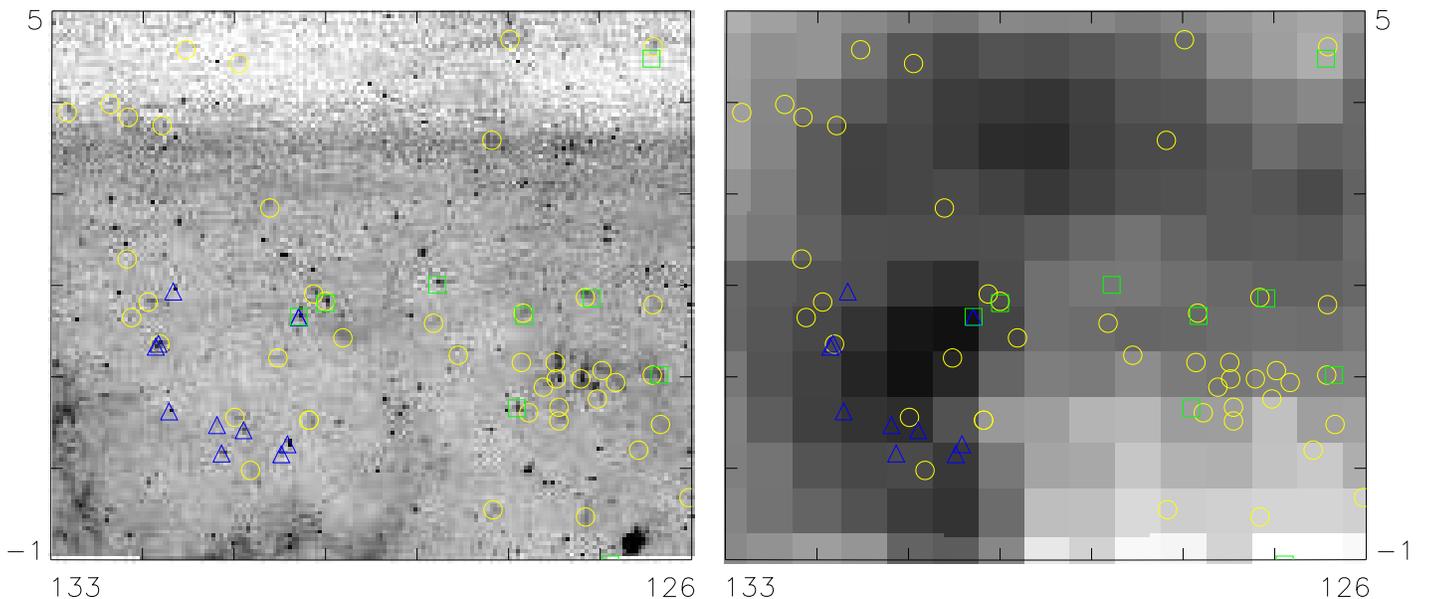}
\caption{(a) MSX 8-$\mu$m deep mosaic image of the center of
  Complex~H.  The displayed region spans Galactic longitudes from
  126\degr\ (right edge of the image) to 133\degr\ (left edge) and
  Galactic latitudes from $-1$\degr\ (bottom) to 5\degr\ (top).  The
  sources we selected for followup observations are marked.  Yellow
  circles represent MSX extended sources, blue triangles represent MSX
  point sources, and green squares represent IRAS sources from
  \citet{ft96}.  (b) \hi\ image of the center of Complex~H from the
  Leiden/Dwingeloo Survey.  The angular coverage is the same as in
  panel (a).  The infrared targets are overplotted as in panel (a).
  It is clear that the location of the \hi\ peak of the HVC lies near
  a minimum of the infrared emission, suggesting that the MSX sources
  are not associated with Complex~H. }
\label{fig:chap3-banda}
\end{figure*}

We then examined the area in question in more detail.  The region of
highest \hi\ column density in the HVC is roughly bounded by $127\degr
< \ell < 133\degr$ and $-1\degr < b < 5\degr$, so for simplicity we
will assume that any star formation that has taken place in Complex~H
occurred in this area.  The MSX band A (8~$\mu$m) deep mosaic image of
this region is shown in Figure \ref{fig:chap3-banda}, along with \hi\
data from the Leiden/Dwingeloo Survey \citep{dap}.  This MSX
image was extracted from a $10\degr\times10\degr$ product with
36\arcsec\ pixels and 72\arcsec\ resolution, and has sensitivity
limits given in \S \ref{chap3-data}.

We selected 43 of the brightest extended sources for follow-up
observations based on a mosaic constructed of high-resolution
(6\arcsec\ pixels, 20\arcsec\ resolution) MSX images covering the
region displayed in Figure \ref{fig:chap3-banda}.  It is clear from a
casual inspection of the figure that the brightest infrared sources in
this part of the sky are concentrated well away from the \hi\ peak of
Complex~H.  We therefore added 10 bright point sources located near
the core of the HVC to our target list.  Finally, we included seven
IRAS sources (many of which were expected to be Galactic star-forming
regions) from the compilation of \citet{ft96} in order to make sure
that they were actually Milky Way objects and to confirm that our
observing strategy could detect star-forming regions.

\subsection{CO Observations}

We observed our targets with the NRAO/UASO 12~m reflector on the days
of 2000 May 5 -- 9.  At the wavelength of the CO \jone\ line, the
telescope has a beam width of 55\arcsec.  We used the millimeter
autocorrelator with a 300 MHz bandwidth and 0.2541 \kms\
resolution and two polarizations.  System temperatures were between
300 and 500 K for most of our observations, and reached as high as
1000 K for a few scans as the HVC approached the horizon.  Typical
integration times were $8-12$ minutes per position, for a total
observing time of $40-60$ minutes on each source.  We observed in
relative position--switching mode with beam throws of a few arcminutes
and used the online vane calibration to set the antenna temperature
scale.  The telescope pointing and focus were checked approximately
every six hours, or after sunrise and sunset.  The data were reduced
in COMB by subtracting a linear baseline from each spectrum and then
combining the two polarizations.  For the extended sources, we
observed a five point cross pattern around each source, with the
spacing of the points chosen to be slightly less than the radius of
the source.  For the point sources, we only observed a single position
since the beam size of the 12~m telescope is larger than the
resolution of the MSX data.

\subsection{Millimeter Search Results}
\label{chap3-results}

We observed the 60 targets shown in Figure \ref{fig:chap3-banda} and
examined the spectra for emission or absorption (emission in the off
position) features in the velocity range $-300 \kmsm < V_{LSR} < 100
\kmsm$.  Milky Way objects should appear with velocities $V_{LSR} >
-120$ \kms, and we expect Complex~H emission to lie between $-230$
\kms\ and $-170$ \kms\ if any is present (velocities as high as $-120$
\kms\ are also possible, but less likely).  We detected Milky Way CO
lines in either emission or absorption toward all 43 of the MSX
extended sources, nine out of the ten MSX point sources, and all seven
of the \citeauthor{ft96} \hii\ regions, indicating that at least 59 of
our 60 targets are almost certainly Milky Way star-forming regions.
The only object that we did not detect, MSX6C~G131.1875+00.4726, may
be a Milky Way infrared source of some kind that does not contain any
CO, or it could be that the molecular gas is simply offset enough from
the infrared position that it lies outside the beam of the 12~m.  We
did not find any emission lines at velocities below $-102$
\kms,\footnote{Several spectra showed apparently significant features
at lower velocities, but repeat observations failed to confirm the
reality of these lines.} so there is no evidence that any of these
targets are associated with Complex~H.

\begin{figure}[th!]
\epsscale{1.2}
\plotone{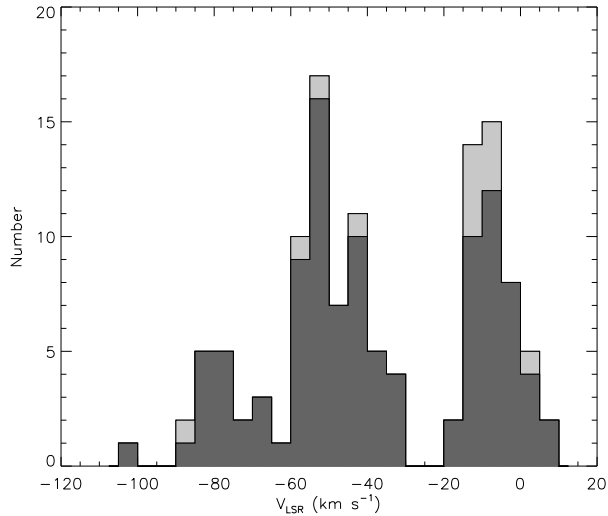}
\caption{Velocity histogram of the Milky Way CO lines we detected.
  Emission lines are shown in dark gray and absorption lines (which
  actually represent emission in the reference position) are shown in
  light gray.}
\label{fig:chap3-velhist}
\end{figure}

In Table \ref{chap3-targets} we list the results of our CO
observations for each source.  Columns (1) and (2) contain the
Galactic coordinates of the targets, column (3) contains the velocity
centroids of the detected lines, column (4) contains the peak observed
brightness temperatures (again, note that negative observed
temperatures indicate the the emission was located in the off
position), column (5) contains the integrated intensity of the line,
and column (6) describes how the targets were selected.  For many
lines of sight multiple emission components were detected, so we
include all lines with an integrated intensity that is significant at
the 5~$\sigma$ level or higher in this table.  Note, however, that we
frequently omit lines at local velocities ($-10 \kmsm \lesssim V_{LSR}
\lesssim 5 \kmsm$) when other components with more negative velocities
are also detected along the same line of sight.  Our compilation is
therefore incomplete at these velocities, but this local material is
not relevant for our purposes.  The distribution of velocities that we
find for the Milky Way clouds we detect is plotted in Figure
\ref{fig:chap3-velhist}.  Distinct peaks in the distribution appear at
velocities of $-10$~\kms\ (local material), $-55$~\kms\ (Perseus
spiral arm), and $-80$~\kms\ (outer spiral arm).

Over the velocity range in which Complex~H emission is most likely
($-230 \kmsm < V_{LSR} < -170 \kmsm$), our spectra reach a typical rms
of 0.05 K for extended sources and 0.13 K for point sources (because
of the shorter integration times).  Assuming a distance of 27 kpc and
a linewidth of 6 channels (1.52 \kms), similar to the linewidths of
the Galactic lines we detected, we translate these rms values to
3~$\sigma$ upper limits on the total molecular mass associated with
each source of $24-130$~M$_{\odot}$, with a median upper limit of
47~M$_{\odot}$.  These masses correspond to typical molecular cores in
Galactic star forming regions, but are much smaller than the masses of
giant molecular clouds.  These mass limits also assume that the Milky
Way value of $X_{CO}$, the CO-H$_{2}$ conversion factor, is
appropriate for Complex~H.  Recent analyses of $X_{CO}$ in other
galaxies suggest that it has at most a weak dependence on metallicity
\citep{walter01,walter02,bolatto03,eros03}, which would validate our
assumption.

In addition to searching for localized CO emission in Complex~H, we
can also combine all of our spectra to search for a very faint,
diffuse component.  Our spectra of 60 sources span 7\degr\ of
longitude and 6\degr\ of latitude, covering a total area of
approximately 0.05 deg$^{2}$ ($\sim10^{4}$~pc$^{2}$ at a distance of
27 kpc) with a net integration time of 125~ks.  The average spectrum
over all of our observed positions has an rms noise level of 5~mK,
which (again assuming the Galactic value of $X_{CO}$ and a linewidth
of 1.52 \kms) yields a 3~$\sigma$ upper limit of $\sim1000$~M$_{\odot}$
on the diffuse molecular content of the HVC.

\subsection{Limits on the Star Formation Rate in Complex~H}
\label{chap3-sflimits}

If we interpret these results as implying that there are currently no
OB associations comparable to or larger than N81 in Complex~H, then
the observed OB star population of N81 suggests that the most massive
star forming region in the HVC must have a star formation rate of less
than $5 \times 10^{-4}$~M$_{\odot}$~yr$^{-1}$.  This limit holds as
long as the distance to Complex~H is less than $\sim75$~kpc (the
largest distance at which N81 would be visible in MSX images).  Since
it is unlikely that Complex~H contains multiple \hii\ regions just
below this limit, this number also represents a plausible upper limit
for the star formation rate of the entire cloud.

An alternate method to estimate the star formation that could be
ongoing in Complex~H without being detected in our data is to utilize
the observed correlations between infrared and radio continuum flux
densities and star formation rates in other galaxies.  \citet{cg01}
showed that \hii\ regions in the Milky Way have a mean ratio of
8~$\mu$m MSX emission to 843~MHz radio continuum emission of $R = 27
\pm 3$.  \citet*{cssg03} suggested that $R$ values of $\approx15$
might be more appropriate for the LMC (and therefore for other dwarfs
as well), but to place the most generous limits we will use the Milky
Way value.  It is well-known that the radio continuum luminosities of
galaxies provide a good indicator of their star formation rates.
\citet{murgia} found that the star formation rate surface density (for
stars more massive than 5~M$_{\odot}$) is $8.0 \times
10^{-4}~B_{1.4}$~M$_{\odot}$~yr$^{-1}$~kpc$^{-2}$, where $B_{1.4}$ is
the observed flux density at 1.4~GHz in mJy/bm.  Since the MSX flux
limit is 100~mJy, our nondetection of any infrared sources associated
with Complex~H implies an upper limit to the 843~MHz radio flux
density of any Complex~H \hii\ regions of 3.7~mJy.
\citet{filipovic98} found approximately flat spectral indices for the
radio continuum emission from LMC and SMC \hii\ regions, so we assume
that the 1.4~GHz flux density of \hii\ regions in Complex~H (if there
are any) is also less than 3.7~mJy.  The \citeauthor{murgia} star
formation rate prescription then indicates that the star formation
surface density in Complex~H star-forming regions must be $\lesssim
3.0 \times 10^{-3}$~M$_{\odot}$~yr$^{-1}$~kpc$^{-2}$.  The 20\arcsec\
MSX beam subtends an area of $5.4 \times 10^{-6} \mbox{ } (d/27\mbox{
kpc})^{2}$~kpc$^{2}$, so the total star formation rate in any single
Complex~H \hii\ region (after accounting for the contribution of
low-mass stars down to 0.1~M$_{\odot}$ with a Salpeter initial mass
function) has an upper limit of $\sim1 \times 10^{-7} \mbox{ }
(d/27\mbox{ kpc})^{2}$~M$_{\odot}$~yr$^{-1}$.  This calculation is
much more restrictive than the one given in the previous paragraph,
but without further knowledge of the infrared and radio properties of
star-forming regions in very low-mass dwarf galaxies, it is difficult
to assess which one provides a more meaningful limit.

By comparison, known Local Group dIrrs have star formation rates
ranging from $1 \times 10^{-4}$~M$_{\odot}$~yr$^{-1}$ for the
Sagittarius dwarf irregular galaxy, with an absolute magnitude of
$M_{V} = -12.3$, up to nearly 1~M$_{\odot}$~yr$^{-1}$ for IC~10, with
an absolute magnitude of $M_{V} = -15.7$ \citep[][and references
therein]{mateo98}.  The most similar dwarfs to Complex~H in terms of
\hi\ mass have star formation rates of a few times
$10^{-4}$~M$_{\odot}$~yr$^{-1}$.  We conclude that the present-day
star formation rate in Complex~H is certainly no higher than that of
other dwarf irregulars with similar masses, and there is some basis
for arguing that it is substantially lower.

\section{SEARCH FOR A DISTANT STELLAR POPULATION BEHIND THE MILKY WAY}
\label{chap3-2mass}

Dwarf galaxies, whether they contain significant amounts of gas or
not, usually have a substantial fraction of their stellar mass locked
up in an old, metal-poor stellar population.  Therefore, even though
Complex~H lacks appreciable amounts of massive star formation, we must
also search for an evolved stellar population associated with the HVC.
The most luminous stars in such a population are on the red giant
branch (RGB) and the asymptotic giant branch (AGB).  These stars can
be easily recognized by their characteristic distribution in optical
and near-infrared color-magnitude diagrams (CMDs).  In the case of
Complex~H, optical data are largely useless because of the very high
foreground density of Milky Way stars combined with several magnitudes
of extinction.  Near-infrared observations, in contrast, offer several
strong advantages: much less extinction, lower foreground levels
(since most stellar spectra peak at shorter wavelengths), and the
enhanced brightness of RGB and AGB stars relative to other types of
stars.  In this section, we describe our use of 2MASS data to search
for evidence of an ancient stellar population in Complex~H.

\subsection{2MASS Data}
\label{chap3-2massdata}

The 2MASS project surveyed the entire sky in three near--infrared
bands ($J$, $H$, and $K_{\mbox{s}}$).  2MASS images have typical
seeing of about 3\arcsec, and the data are complete down to
10~$\sigma$ limiting magnitudes of $J=15.8$, $H=15.1$, and
$K_{\mbox{s}}=14.3$ (2MASS Explanatory
Supplement.\footnote{http://www.ipac.caltech.edu/2mass/releases/allsky/doc/explsup.html})
In the near--infrared, the red giant branch extends up to
$K_{\mbox{s}} \approx -6.2$ at a color of $J-K_{\mbox{s}} \approx 1$
\citep{nw00}, so a population of evolved stars is visible in the 2MASS
dataset out to a distance modulus of $m - M \approx 20$ ($d = 100$
kpc).  The 2MASS Point Source Catalog \citep{2masspsc} includes
photometry and astrometry for $\sim4.7 \times 10^{8}$ objects.

\subsection{Search Technique}
\label{chap3-searchtech}

Using the 2MASS Point Source Catalog, we can construct
color--magnitude diagrams (CMDs) of the center of Complex~H.  We
search for a population of evolved stars behind the Milky Way by
comparing these CMDs to those of nearby regions (e.g., at the same
Galactic latitude but $\sim 10\degr$ away in longitude).  If the
population is relatively massive (comparable to the Sagittarius dSph,
for example), the RGB will be visually obvious in the CMD without even
attempting to enhance the signal by statistically removing foreground
stars.  Low-mass dwarfs, on the other hand, can be difficult to
detect.

\begin{figure*}[th]
\epsscale{0.85}
\plotone{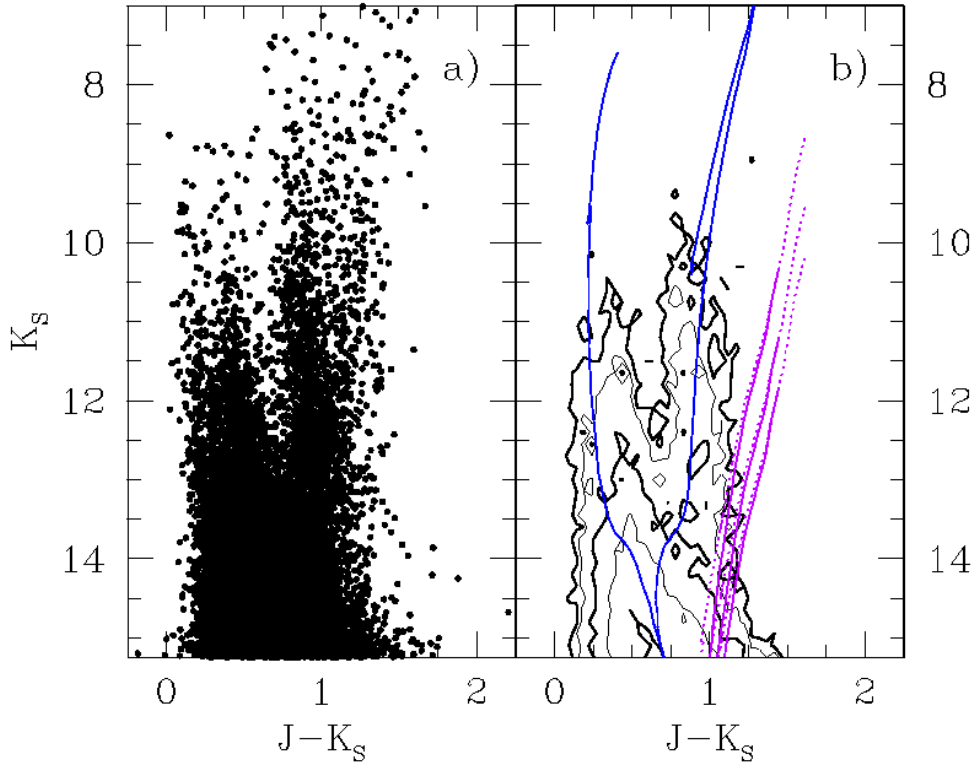}
\caption{(a) 2MASS color-magnitude diagram of the center of
    Complex~H.  (b) 2MASS CMD (contours) with theoretical stellar
    isochrones overlaid.  The stellar density is contoured with
    logarithmically spaced isopleths between 4 and 144 stars per 0.1
    mag$^{2}$.  The isochrones are based on the Padua set of stellar
    models \citep{girardi00} and show the expected locations of
    Galactic and putative dwarf galaxy stellar populations in the
    diagram.  Solar metallicity tracks ($Z = 0.019$) are shown in
    blue, and low-metallicity tracks ($Z = 0.004$) are displayed in
    purple.  For display purposes, the high-metallicity tracks are
    shown for the single distance of 2~kpc (the approximate distance
    to the nearest spiral arm in this sightline; \citealt*{gbc02}).
    The ages for these tracks are 100~Myr, to show the approximate
    location of the zero-age main sequence (the plume of stars with
    $J-K_{\mbox{s}} \lesssim 0.75$), and 10~Gyr, to give an
    approximate trace of the red giant branch populations (the redder
    plume with $J-K_{\mbox{s}} \approx 1.0$).  A reddening equivalent
    to E(B-V) = 0.6 has been applied to the isochrones representing
    these relatively nearby populations.  The primary features visible
    in the CMD are clearly caused by Milky Way main sequence and red
    giant stars present over a wide range of distances.  The purple
    tracks show a relatively metal-rich ($[Fe/H] \approx -0.7$) dwarf
    galaxy population, with solid lines representing stars up to the
    tip of the red giant branch, and dotted lines to show the
    horizontal branch and AGB evolution.  Three distances have been
    chosen for display, corresponding to the likely distance of
    Complex~H: 18, 27, and 36~kpc.  All three tracks have been plotted
    with the equivalent of 0.8~mag of foreground reddening, and an age
    of 4~Gyr. }
\label{fig:chap3-cmd}
\end{figure*}

In Figure \ref{fig:chap3-cmd} we display the $K_{\mbox{s}}$,
$J-K_{\mbox{s}}$ CMD for a 1~deg$^{2}$ region centered on the peak of
the HI distribution of Complex~H ($\ell = 131\degr, b = 1\degr$).  The
two strong plumes of stars extending upward around $J-K_{\mbox{s}} =
0.4$ and $J-K_{\mbox{s}} = 1.0$ correspond to foreground main sequence
stars and a mixture of Milky Way red clump and red giant branch stars,
respectively.  As shown in Figure \ref{fig:chap3-cmd}\emph{b}, a red
giant population associated with Complex H would appear as a sequence
of stars following the purple tracks up and to the right from the
right side of the foreground giant branch \citep[e.g.,][]{cole01}.
Age differences only have a very small effect on the color of the
dwarf galaxy giant branches, while a decrease in reddening or
metallicity would move the tracks to the blue, increasing the
possibility of confusion with the plume of foreground K giants.  No
feature that is immediately identifiable with a dwarf galaxy
population is apparent, although there is a small scattering of stars
fainter and redder than $(J-K_{\mbox{s}}, K_{\mbox{s}}) = (1.2, 13)$
that could conceivably be representatives of such a population.
However, these stars could equally well be differentially reddened
foreground K giants at distances of a few kiloparsecs.

\begin{figure*}[th]
\epsscale{1.1}
\plotone{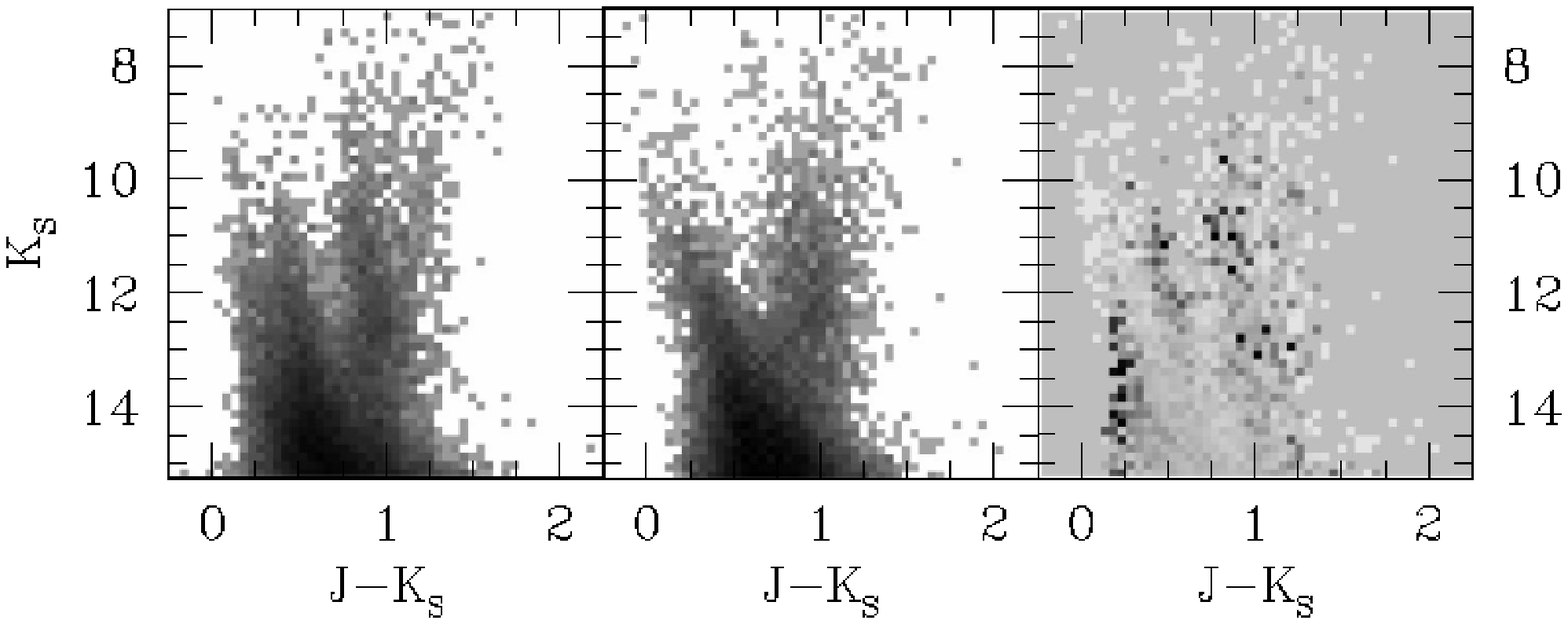}
\caption{Hess diagrams of the 2MASS data in the direction of
    Complex~H (left panel) and the control field (middle panel).  The
    grayscale ranges from 0 counts per cell (white) to 125 counts per
    cell (black).  The difference between the two Hess diagrams, after
    adjusting the reddening of the control field to match that of
    Complex~H, is displayed in the right panel.  The cells in the
    residual plot represent counts in the Complex~H Hess diagram minus
    counts in the control field Hess diagram, as a fraction of the
    counts in the control field Hess diagram.  The mean difference
    level is about 7\%, and the grayscale ranges from $-1$\% (white)
    to 10\% (black).  Although the residuals shows that the
    populations of the two fields are not identical, there is no
    coherent structure at the location where a red giant population at
    $d \sim 27$~kpc would be found (see Figure \ref{fig:chap3-cmd}).}
\label{fig:chap3-hess}
\end{figure*}

To test the significance of this scattered population of faint red
stars, we compared the Complex~H field to a nearby control field.
Because the sightline through the Galactic plane is subject to strong
differential reddening, a random choice of control field can yield
misleading results.  We used MSX band A images to guide us to an
appropriate control field, operating under the assumption that areas
with low levels of emission from warm dust (which is visible at
8~$\mu$m) would also have low levels of differential reddening.  On
this basis, we selected a 1~deg$^{2}$ control field located at $\ell =
141\degr, b = 2.2\degr$.  We show grayscale Hess diagrams of the
Complex~H and control fields in Figure \ref{fig:chap3-hess} (left and
center panels, respectively).  As measured by the width of the
foreground red giant branch and the number of highly reddened
outliers, the two fields are indeed subject to similar amounts of
differential reddening.  We control for any difference in the mean
reddening by measuring the color of the main-sequence stars in each
field at $13 < K_{\mbox{s}} < 14$.  The measured colors of
$J-K_{\mbox{s}} = 0.52$ (Complex~H) and $J-K_{\mbox{s}} = 0.67$
(control field) imply a difference in E(B-V) of 0.27~mag.  We
dereddened the control field by this amount and subtracted the two
Hess diagrams, producing the difference image shown in the right panel
of Figure \ref{fig:chap3-hess}.  No coherent structure corresponding
to a dwarf galaxy RGB is evident.

\subsection{Detection Limits}
\label{chap3-detectlimits}

In order to quantify the limits that the 2MASS data place on the
stellar content of the HVC, we considered the detectability of various
stellar populations in the 2MASS observations of Complex~H.  We first
determined the surface brightness limit of the 2MASS point source
catalog for the detection of resolved systems at various distances
\emph{along this particular sightline}.  To do so, we calculated the
number of excess stars necessary to constitute a 3~$\sigma$ detection
of a dwarf galaxy in the Complex~H field that is not present in the
control field.  The star counts are related to $V$-band surface
brightnesses using empirical $K_{\mbox{s}}$-band luminosity functions
and starcount-surface brightness relations for dwarf spheroidals taken
from the literature.  Owing to the survey's rather shallow depth, we
restrict our attention to the part of the CMD that we would expect to
find populated with the red giant stars of a nearby dwarf galaxy.
This isolates the relevant stars and suppresses noise from blue stars
that are certain to be foreground objects.  In raw terms, the
detection limits on a stellar population are set by Poisson
statistics: a galaxy of given surface brightness is expected to
contain a certain number of RGB and AGB stars detectable by 2MASS.  By
comparing the control field $+$ model dwarf galaxy starcounts to the
Complex~H starcounts, we can put a lower (bright) limit on the surface
brightness of any possibly present dwarf.

Broadly speaking, there are two generic categories of dwarf galaxies:
those with substantial intermediate-age populations (e.g., Fornax,
Leo~I, and the Sagittarius dSph), and those without (e.g., Sculptor,
Ursa~Minor, and Draco).  Therefore we used two different sets of
luminosity functions, color cuts, and starcount-surface brightness
relations in our analysis.  We modeled both types of dwarfs with a
luminosity function (LF) of the same form, $N(K) \propto 10^{K/4}$
\citep{davidge00a,davidge00b}.  The luminosity functions are taken to
be identical for magnitudes below the RGB tip ($K_{\mbox{s}} \approx
-6.2$); brighter than this limit, the LF is assumed to drop by a
factor of 3 in the mixed-age population, giving it a significant
population of AGB stars.  This model contains roughly equal fractions
of old and intermediate-age stars, similar to Fornax
\citep*{shb00}.  By contrast, the purely old population suffers a
factor of 10 discontinuity at the tip of the RGB and therefore
contains virtually no bright AGB stars.

The normalization factors and conversion to starcounts were taken from
\citet{ih95}, with our purely old galaxy typified by the Sculptor
dSph, and using the Fornax dSph as our template for an
intermediate-age dwarf.  In making the conversion between galaxy
surface brightness and predicted 2MASS starcounts it is important to
consider the effect of incompleteness on the faint end of the
luminosity function.  We modeled incompleteness by multiplying the
synthetic luminosity functions by a function proportional to tanh($K$)
to generate our starcount predictions.

\begin{figure}[th]
\epsscale{1.25}
\plotone{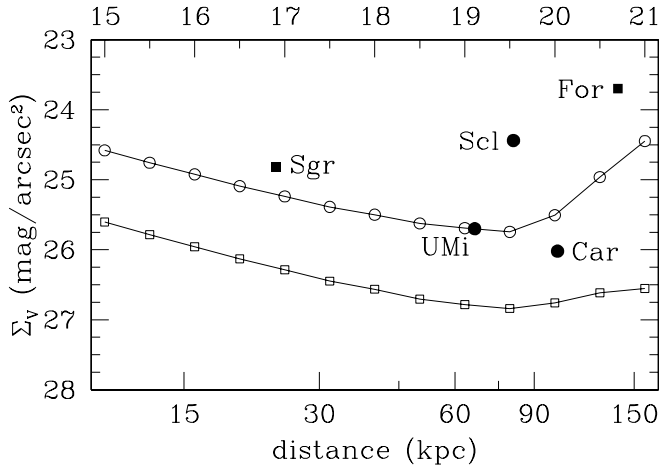}
\caption{3~$\sigma$ detection limits for old and intermediate age
    stellar populations at the position of Complex~H as a function of
    distance and surface brightness.  The scale along the top of the
    figure shows distance modulus (in magnitudes).  The open circles
    represent the detection limits for a dwarf containing only an old
    population of stars, and the open squares represent the detection
    limits if an intermediate-age population is present as well.  For
    reference, the locations of a subset of known Milky Way satellites
    are shown with solid points --- circles for the purely old systems
    (compare to the brighter detection limit), and squares for the
    mixed-age systems (compare to the fainter detection limit).  Note
    that the surface brightnesses plotted for these galaxies represent
    the mean level over 1~deg$^{2}$, not the central surface
    brightness.  At a distance of 27~kpc, the surface brightness
    limits are $\mu_{V} = 25.25$ \surfb\ (old) and $\mu_{V} = 26.25$
    \surfb\ (intermediate), so the only known Local Group dwarfs that
    would not be detected at this position for reasonable distances
    are the very lowest-mass dwarf spheroidals (Carina, Sextans, Ursa
    Minor, Ursa Major, and And~IX).}
\label{fig:chap3-detlimits}
\end{figure}

We then computed lower limits to the surface brightness of a putative
dwarf galaxy associated with Complex~H for the adopted reddening (see
\S \ref{chap3-searchtech}) and for distances ranging from
$10-160$~kpc.  The results are shown in Figure
\ref{fig:chap3-detlimits}, where the open circles represent our
3~$\sigma$ detection limits for a dwarf galaxy dominated by an old
stellar population at this position, and the open squares represent
the same limit for a galaxy that also contains an intermediate-age
component.  Note that the shape of the detectability curves is very
similar to those calculated by \citet{willman02a} for the \emph{Sloan
Digital Sky Survey} (SDSS), but the SDSS data probe $\sim2$~mag
deeper.

These calculations place a firm lower limit on the surface brightness
of the old stellar content of Complex~H of $\mu_{V} \ge 25.25$~\surfb.
Spread over our 1~deg$^{2}$ search area, the corresponding stellar
luminosity is $6 \times 10^{5}$~L$_{\odot}$, implying an absolute
upper limit to the stellar mass of Complex~H of $1.2 \times
10^{6}$~M$_{\odot}$ for a typical $V$-band stellar mass-to-light ratio
of 2.  For a population covering a smaller area, the stellar mass
limit is correspondingly lower.  If Complex~H contains an
intermediate-age population, then the surface brightness limit is
$\mu_{V} \ge 26.25$~\surfb, at least 1~\surfb\ fainter than that of
the Sgr dSph and 2.5~\surfb\ fainter than Fornax.  No known dwarf
irregular galaxies have such low surface brightnesses, and only
$\sim1/4$ of the currently known Local Group dwarf spheroidals are so
faint.  In the case of a purely old population of stars, a galaxy
comparable to Carina or Ursa Minor would be marginally detectable in
the 2MASS data.  Extremely low-mass dSphs such as Ursa Major and
And~IX could therefore hide in the Galactic foreground below our
detection limits, but there are no known dwarf irregular galaxies that
contain as much gas as Complex~H does and would not be detectable in
2MASS at a distance of 27~kpc.  We conclude that Complex~H would have
to be a unique hybrid dwarf galaxy, with a large gas mass and low
surface brightness, in order to contain a stellar population that is
not visible in the 2MASS data.

A previous search of 2MASS data for an RGB associated with Complex~H
was carried out by S. Majewski and M. Skrutskie and also did not
detect any evidence for a Complex~H stellar population
\citep{lockman}, consistent with our results.

\section{DISCUSSION}
\label{chap3-discussion}

\subsection{What Is Complex H?}
\label{chap3-whatisit}

W98 presented six plausible explanations of the origin of Complex~H,
ranging from the neutral edge of a superbubble in the outer Galaxy to
a massive, distant intergalactic cloud in the Local Group.  Given the
distance limits discussed in \S \ref{chap3-intro}, the remaining
possibilities are (1) Complex~H was produced by a Galactic fountain,
(2) Complex~H is an infalling extragalactic cloud, and (3) Complex~H
is a nearby dwarf galaxy.  A Galactic fountain origin for Complex~H
cannot be ruled out entirely without absorption-line measurements, but
the low metallicities of other HVCs
\citep[e.g.,][]{lu94a,lu94b,wakker99,w01,richter01,sembach02,tripp03}
and its large mass and distance suggest that this possibility is
unlikely.  To the remaining two possibilities, we add a third related,
but distinct scenario: that Complex H represents gas that has been
tidally stripped from a Milky Way satellite galaxy.

A stripped gas origin for Complex H also appears improbable for two
reasons.  First, the morphology of the HVC (Figure
\ref{fig:chap3-ldsmap}) does not appear consistent with that of a
tidal stream.  Rather than a highly elongated, symmetric structure
(perhaps with sizeable remnant in the middle of it, if the progenitor
has not been entirely destroyed), Complex H is asymmetric, with one
end much denser than the other, and only moderately elongated (axis
ratio of less than 2:1).  The only way that Complex H could be a
classical tidal stream is if the direction of the stream is along the
line of sight.  This orientation would seem to be at odds with the
orbit discussed by \citet{lockman}, which is roughly circular and
inclined by 45\degr\ with respect to the Galactic plane, indicating
that the direction of motion of Complex H is primarily perpendicular
to the line of sight.  The second difficulty with the tidal
explanation is the lack of a credible progenitor object.  Unlike the
one HVC feature that is known to result from stripping of a dwarf
galaxy, the Magellanic Stream, there are no known dwarf galaxies
connected to any part of Complex H.  Depending on the initial
characteristics of this putative object (stellar mass, and relative
extents of the gas and stars), one might expect a significant fraction
of the stellar component to still be present at the location of the
gas; the results of this paper rule such a possibility out.
Therefore, the original stellar component must either be spread out in
a stream along the orbit or exist as a bound object elsewhere on the
orbit.

Complex~H does not appear to host either significant amounts of recent
star formation or a substantial old stellar population, making it
difficult to distinguish between the two remaining possibilities.  If
Complex~H is associated with a dwarf galaxy, it must be a rather faint
dwarf to have escaped detection.  The combination of a low total
luminosity, small numbers of evolved stars, low levels of ongoing star
formation, and a large gas mass would make this dwarf galaxy unique in
the Local Group.

On the other hand, if Complex~H is \emph{not} a dwarf galaxy, we would
be forced to conclude by default that it is an infalling extragalactic
cloud that is in the process of being accreted by the Milky Way.  In
this case, we must understand why Complex~H has failed to form any
stars despite the $\gtrsim 10^{7}$~M$_{\odot}$ of \hi\ that it
contains.  What makes Complex~H different from other gas clouds of
similar masses?  Did it only acquire its gas recently, not allowing
time for star formation?  Is its internal pressure too low for
molecular clouds to form \citep{bros04}?  Higher resolution \hi\
observations of Complex~H support the latter possibility, indicating
peak \hi\ column densities of $\sim2.5 \times 10^{20}$~cm$^{-2}$ and
number densities of a few particles per cm$^{3}$
\citep*{ws91,wvs91,lockman}.  However, these measurements do not
address the more fundamental issue of \emph{why} the pressure and
density in such a large cloud of gas are so low.

The key question that must be answered to resolve this issue is
whether Complex~H is surrounded by a dark matter halo, and therefore
represents CDM substructure, or whether it is a cold accretion flow
such as those recently suggested by \citet{bd03}, \citet{keres04}, and
\citet{db04}.  The argument for Complex~H being a dark galaxy relies
on the CDM prediction of large numbers of low-mass dark matter halos,
its large \hi\ mass, and its almost unprecedented \hi-to-stellar mass
ratio.  If the predicted subhalos exist (as we argue is likely; see \S
\ref{chap3-substructure}), then the most massive subhalos should have
retained or accreted some gas.  Complex~H seems to have many of the
expected properties of one of these objects.  The only hole in this
case is the kinematics of the cloud.  As described in \S
\ref{chap3-intro}, a cloud as massive as Complex~H should have a
rotation velocity of $\sim30$~\kms, implying a velocity gradient as
large as 60~\kms, or a velocity dispersion of similar magnitude.  The
observed gradient across the HVC is closer to 20~\kms, and the
velocity dispersion is about 12~\kms, much smaller than expected.
These values can still be compatible with a large total mass for
Complex~H if (1) we are seeing the rotation close to face-on, and
therefore underestimating its amplitude, or (2) the \hi\ is confined
to the central regions of the dark matter halo, so that the kinematics
do not reveal the full mass of the cloud.  It is worth noting that
other Local Group dwarfs with similar \hi\ masses to Complex~H (e.g.,
Sextans~B and IC~1613) also have fairly low rotation amplitudes
($\sim10$~\kms) and velocity dispersions.

Because this interpretation requires somewhat special circumstances,
the alternative possibility that Complex~H represents an instance of
``cold mode'' gas accretion onto the Milky Way (as opposed to the
standard ``hot mode'', in which gas is shock heated to very high
temperatures and ionized at large radii before accreting onto a galaxy
and later cooling) must be considered as well.  Complex~H would be the
first observed example of this process, which has heretofore only been
seen in hydrodynamic simulations.  This would constitute strong
evidence that cold accretion actually does play an important role in
galaxy formation in the real universe.  In addition to the kinematics,
another piece of evidence in support of this idea is that Complex~H is
located relatively close on the sky to M31, which corresponds to the
direction along the filament from which the Local Group formed
\citep{b99,khkg}.  However, there are also problematic aspects to this
picture, notably that Complex~H cannot be gravitationally bound
without large amounts of dark matter, and therefore would necessarily
be a transient object that formed recently, and also that similar
objects have not been detected around other galaxies even though
surveys with sufficient sensitivity to detect Complex~H analogs have
been carried out.  More detailed observations of the kinematics of
Complex~H may help to determine which of these possibilities is
correct, but the more basic problem that the behavior of gas clouds in
the Local Group is not yet understood \citep{smw02} also must be
addressed.

\subsection{Implications for the Nature of HVCs}
\label{chap3-hvcs}

Complex~H now joins an increasingly long list of other high-velocity
clouds that appear not to contain stars
\citep{sb02,davies02,willman02b,hopp03,siegel05}.  The hypothesis that
HVCs are the \hi\ counterparts of normal low-surface brightness dwarf
galaxies can therefore safely be put to rest.  It remains possible, of
course, that HVCs could host stellar populations whose surface
brightness is significantly lower than that of any currently known
galaxy, but more likely, HVCs simply do not contain any stars.

A number of competing models to explain the origins of HVCs have been
proposed in recent years: (1) HVCs are distant ($d \gtrsim 500$~kpc),
massive clouds that are the left-over building blocks of the Local
Group \citep{b99}, (2) HVCs are small, nearby ($d \lesssim 50$~kpc)
clouds that represent tidal debris from destroyed dwarf galaxies, and
(3) HVCs are low-mass clouds at intermediate distances ($d \lesssim
150$~kpc) that are cooling and condensing out of the halo of hot gas
that surrounds the Milky Way \citep{mb05}.  Only in the first model
would HVCs be expected to be associated with dwarf galaxies.  The
absence of stars in HVCs, however, is not necessarily evidence against
this model because of the many possible mechanisms for suppressing
star formation in low-mass objects (see \S \ref{chap3-intro}).  Our
results therefore do not provide us with significant new leverage on
the nature of HVCs.  Still, it may be worth noting that Complex~H is
cooler ($T \sim 50$ K; \citealt{wvs91}) and possibly more massive than
the clouds expected in the \citet{mb05} model.

\subsection{Implications for the Substructure Problem}
\label{chap3-substructure}

According to $\Lambda$CDM numerical simulations, the Local Group
should contain up to $\sim500$ low-mass dark matter halos
\citep{klypin99,moore99}.  Observationally, there are less than 40
known Local Group dwarf galaxies
\citep[e.g.,][]{mateo98,vandenbergh00}.  Possible explanations for
this mismatch are: (1) the simulations are overpredicting the amount
of substructure that should be present, (2) there are many faint Local
Group dwarfs that have not yet been discovered, or (3) $\sim90$~\% of
low-mass halos never form stars.  There are only a few ways to alter
the predictions of the simulations, such as changing the initial power
spectrum of density fluctuations \citep{kl00} or imbuing the dark
matter particles with new properties (e.g., a nonzero self-interaction
cross-section [\citealt{ss00}] or annihilation rate [e.g.,
\citealt{kaplinghat00}]).  Because there is little theoretical or
observational motivation for these ideas at present, much of the
attention has focused on the other two potential solutions of the
substructure problem.

Recent observational work has cast doubt on the idea that hundreds of
undiscovered dwarf galaxies could exist in the Local Group.
\citet{willman04} estimate that at most $\sim9$ galaxies similar to
the population of known Local Group dwarfs may still remain
undiscovered around the Milky Way as a result of extinction and
insufficient sensitivity at large radii.  Our results reported here,
along with the optical studies of other HVCs cited above (\S
\ref{chap3-hvcs}), show that HVCs do not contain dwarf galaxy-like
stellar populations.  The most straightforward interpretation of these
findings is that there are no stars in HVCs; if any stars have formed,
the process must have been extremely inefficient.  We therefore argue
that the substructure problem likely originates in the difficulty that
low-mass dark matter halos experience in becoming dwarf galaxies.  A
number of theoretical ideas support the plausibility of this
hypothesis.  For example, the heating of the intergalactic medium that
occurred during the epoch of reionization may have prevented low-mass
halos from holding on to their gas \citep{bullock00}.  Simple
photoionization caused by the ultraviolet background can also prevent
these halos from forming stars \citep{somerville}.  Tidal stripping
may have removed most of the mass from dwarf galaxies relatively early
in the history of the universe \citep{kgk04}.  Alternatively, the
supernova-driven winds produced by the formation of massive galaxies
at high redshift could have blown out the interstellar medium of their
satellite galaxies \citep{evan00,evan01}.

Observations are also beginning to provide evidence in favor of this
picture as well.  Flux anomalies in multiply-imaged quasars appear to
require that the lensing galaxies contain significant substructure
\citep{ms98,chiba02,mz02,dk02,kd04}.  It is not yet clear whether
these substructures are luminous or dark, but their abundance is
roughly consistent with $\Lambda$CDM predictions.  In the nearby
universe, \citet*{rsb02} recently identified a mysterious object that
may be the first of the posited dark galaxies to be discovered in the
Local Group.  This object is a high-velocity cloud
(HVC~127$-$41$-$330) that is apparently interacting with the Local
Group dwarf galaxy LGS~3 and therefore is located at a distance of
$\sim700$~kpc from the Milky Way.  The velocity field of this HVC
indicates that it is rotating, and the inferred total mass is at least
four times as large as the \hi\ mass.  A preliminary search for stars
in HVC~127$-$41$-$330 did not yield any evidence for a stellar
component \citep{rsb02}, leading us to conclude that this object has
all of the expected characteristics of one of the missing dark matter
halos from CDM simulations \citep*{srb03}.  \citet{thilker04} and
\citet*{wbt05} also argue that some, although not all, of the newly
discovered HVCs around M31 may be dark matter-dominated Local Group
subhalos.

Despite this apparently promising support for the possibility of
inefficient star formation in low-mass objects resolving the
substructure problem, it is worth noting that several extremely faint
and low surface brightness objects have recently been discovered in
the Sloan Digital Sky Survey, suggesting that the census of luminous
dwarfs in the Local Group is not yet complete
\citep{zucker04,willman05,willman05b}.

\subsection{Starless \hi\ Clouds in the Local Group}
\label{chap3-starlessclouds}

Do dark galaxies actually exist?  Blind \hi\ surveys covering large
areas of sky and smaller targeted surveys of nearby galaxy groups have
turned up small numbers of low-mass \hi\ clouds, but deep optical
imaging almost invariably reveals that these objects are associated
with dwarf galaxies \citep*[e.g.,][]{banks,pwl02}.  The only exception
is the object recently announced by \citet{minchin05}, VirgoHI21.
This $10^{8}$~M$_{\odot}$ cloud in the Virgo cluster has no optical
counterpart down to a surface brightness limit of $\mu_{B} =
27.5$~\surfb, which would make it the lowest surface brightness galaxy
known if it does turn out to contain any stars.  VirgoHI21 is
currently the only such starless cloud known outside the Local Group,
and therefore by default represents the closest extragalactic analog
to Complex~H, despite its large mass and location in a cluster
environment.  \citet{minchin05} suggest that previous \hi\ surveys
have not reached low enough column density limits to detect other
similar objects, so more sensitive \hi\ surveys may yet reveal a
larger population of starless \hi\ clouds (however, see
\citealt{zb00,zwaan01}).

Based on a thermal instability model of gas clouds embedded in dark
matter halos, \citet{tw05} conclude that objects with \hi\ masses
comparable to Complex~H are likely ($> 50\%$ probability) to form
stars.  However, the minimum self-regulating star formation rate that
they predict for a $10^{7}$~M$_{\odot}$ cloud is
$\sim10^{-6}$~M$_{\odot}$~yr$^{-1}$, significantly below the level
that we would be able to detect; this model therefore does not require
Complex~H to be completely starless.

If Complex~H and HVC~127$-$41$-$330 indeed lack any stellar component,
then these clouds, along with Wright's Cloud \citep{wright74,wright79}
and Davies' Cloud \citep{davies75}, may be the most massive examples
of a population of starless clouds of gas in the Local Group.  It is
tempting to associate these objects with some of the dark matter
substructures predicted by CDM, but further kinematical studies will
be necessary to determine whether Complex~H, Wright's Cloud, and
Davies' Cloud are in fact dark matter--dominated.  The association of
these objects with dark matter halos would provide strong evidence
that the CDM substructure predictions are correct and that many more
low-mass halos do exist in the Local Group, although the lower-mass
halos may contain much less neutral gas (or none at all), making them
more difficult to detect.  Future multiwavelength investigations of
these massive \hi\ clouds to ascertain how they are distinct from
similar-sized clouds that formed dwarf galaxies will help to
illuminate the process of galaxy formation at both high and low
redshift.

\section{Conclusions}
\label{chap3-conclusion}

We have sought infrared and radio evidence of a dwarf galaxy
associated with the massive, nearby HVC Complex~H.  \citet{lockman}
showed that Complex~H is most likely located slightly beyond the edge
of the Milky Way disk at a distance of $27 \pm 9$~kpc from the Sun,
consistent with the nondetection of high-velocity absorption lines in
the spectra of distant OB stars \citep{w98}.  Mid-infrared
observations from the MSX satellite reveal many star-forming regions
in the Galactic plane near the position of Complex~H.  We used CO
observations to measure the velocities of 59 such sources and
determined that they are clearly Milky Way objects.  Only one of the
sources was not detected in CO and therefore has an unknown origin,
but it is plausible that our observations happened to miss the Milky
Way CO cloud near this line of sight.  Under the assumptions that
Complex~H lies at the distance of 27 kpc suggested by \citet{lockman},
and that the Galactic CO-H$_{2}$ conversion factor applies to HVCs, we
placed 3~$\sigma$ upper limits of $24-130$~M$_{\odot}$ on the amount
of molecular gas that could be present in Complex~H at the position of
each of these targets.  These limits allow us to rule out the
possibility that any of the infrared sources are associated with
typical star-forming molecular cores or giant molecular clouds in
Complex~H.  We also derived an upper limit of $\sim1000$~M$_{\odot}$
on the mass of diffuse molecular gas that Complex~H could contain.
Assuming that there are no massive star-forming regions comparable to
those in other Local Group dwarfs currently present in Complex~H, as
these observations suggest, we placed upper limits on the star
formation rate in the HVC of $5 \times 10^{-4}$~M$_{\odot}$~yr$^{-1}$.
An alternative calculation based on the observed correlations between
infrared emission, radio continuum emission, and star formation
surface density in galaxies and star-forming regions produces the much
stricter limit of $1 \times 10^{-7} \mbox{ } (d/27\mbox{
kpc})^{2}$~M$_{\odot}$~yr$^{-1}$.  If the actual star formation rate
in the HVC is close to the higher of these limits, Complex~H would
rank near the bottom of the list of Local Group dwarf irregulars in
star formation despite its large reservoir of atomic gas.

We also used 2MASS data to constrain the possible intermediate and old
stellar populations within Complex~H.  Near-infrared color-magnitude
diagrams of the core of the HVC do not contain any detectable excess
of RGB or AGB stars compared to a nearby control field.  We built
model stellar components for dwarf galaxies with purely old and
old+intermediate populations and added them to the color-magnitude
diagram of the control field to test their detectability.  We
determined that the nondetection of Complex~H in 2MASS places a lower
limit of $\mu_{V} = 25.25$~\surfb\ on its intrinsic surface brightness
(assuming a distance of 27~kpc), indicating that most of the known
dwarf galaxies in the Local Group (and all of the known dwarfs with
gas) could be detected at the position of Complex~H.  The
corresponding upper limit to the stellar mass of Complex~H is
$10^{6}$~M$_{\odot}$, assuming a $V$-band stellar mass-to-light ratio
of 2.  If a significant intermediate-age population is present as
well, the surface brightness limit improves to 26.25~\surfb, and the
stellar mass limit is even lower.

While we cannot entirely rule out the existence of a dwarf galaxy
associated with Complex~H, these results demonstrate that the galaxy
must have a unique set of properties and a rather low surface
brightness if it exists.  There are no known dwarfs that have the
combined characteristics of a large gas mass, very low current rates
of star formation, and a very small intermediate and old stellar
population.  If Complex~H does contain any stars, it almost certainly
has the largest $M_{\mbox{\small{\hi}}}/L_{B}$ and one of the smallest
populations of evolved stars in the Local Group.

We presented arguments that Complex~H is either a dark galaxy in the
Local Group, or an example of a cold accretion flow onto the Milky
Way.  In either case, this HVC is a unique object whose existence has
important implications for our understanding of galaxy formation, and
further observations and modeling will be needed to determine the true
nature of Complex~H.  If Complex~H is a dark galaxy, and the Cold Dark
Matter prediction of hundreds of low-mass dark matter halos in the
Local Group is correct, then our results provide observational
evidence that these halos may be unable to form stars.  Some of the
most massive subhalos, such as Complex~H and the dark
matter--dominated HVC described by \citet{rsb02} may have accumulated
enough gas to be detected as high-velocity clouds, but the bulk of the
population could remain entirely dark.

\acknowledgments{This research was partially supported by NSF grants
AST-9981308 and AST-0228963.  The authors acknowledge the referee,
Steven Majewski, for suggestions that helped to clarify and improve
the paper.  JDS would like to thank Tam Helfer for her introduction to
COMB and observing with the 12~m.  We also acknowledge the assistance
of telescope operators Duane Clark and Kevin Long and express
appreciation for their ability to keep the telescope running.  JDS
also thanks Tim Robishaw for his usual assistance with IDL-related
matters.  MC thanks NASA for supporting his participation in this work
under LTSA grant NAG5-7936 with UC Berkeley.  This work makes use of
data products from the Midcourse Space Experiment.  Processing of the
data was funded by the Ballistic Missile Defense Organization with
additional support from the NASA Office of Space Science.  This
publication also makes use of data products from the Two Micron All
Sky Survey, which is a joint project of the University of
Massachusetts and the Infrared Processing and Analysis
Center/California Institute of Technology, funded by the National
Aeronautics and Space Administration (NASA) and the National Science
Foundation.  This research has also made use of NASA's Astrophysics
Data System Bibliographic Services. }

\bigskip
\bigskip

\LongTables
\begin{deluxetable}{c c c c c c}
\tabletypesize{\scriptsize}
\tablewidth{0pt}
\tablecolumns{6}
\tablecaption{Milky Way CO Lines Detected Toward Mid-Infrared Sources}
\tablehead{
\colhead{$\ell$} & \colhead{$b$} & \colhead{$V_{LSR}$ (\kms)} &
\colhead{$T_{b}$ (K)} & \colhead{$I_{CO}$ (K~km~s$^{-1}$)} &
\colhead{Source} \\
\colhead{(1)} & \colhead{(2)} & \colhead{(3)} &
\colhead{(4)} & \colhead{(5)} & \colhead{(6)} }

\startdata 
126.03\phn &  $-$0.32\phn & \phn$-$47.0       & \phs\phn5.84 & \phs\phn8.16 & MSXE \\
126.34\phn & \phs0.48\phn & \phs\phn\phn1.9   & \phs\phn4.89 & \phs\phn5.35 & MSXE \\
           &              & \phn\phn$-$4.6    & \phs\phn1.49 & \phs\phn1.89 &      \\
           &              & \phn$-$12.8       & \phs\phn3.61 & \phs10.66    &      \\
           &              & \phn$-$45.3       & \phs\phn1.65 & \phs\phn2.23 &      \\
126.41\phn & \phs4.61\phn & \phs\phn\phn3.8   & \phs\phn2.15 & \phs\phn2.78 & MSXE \\
           &              & \phn\phn$-$6.5    & \phs\phn0.97 & \phs\phn2.01 &      \\
           &              & \phn\phn$-$9.8    & \phs\phn0.51 & \phs\phn0.50 &      \\
126.42\phn & \phs1.79\phn & \phn$-$11.5       & \phs\phn0.52 & \phs\phn0.41 & MSXE \\
           &              & \phn$-$46.2       & \phs\phn0.41 & \phs\phn0.59 &      \\ 
126.43\phn & \phs1.02\phn & \phs\phn\phn1.4   & \phn$-$0.65  & \phn$-$1.30  & MSXE \\
           &              & \phn$-$67.2       & \phs\phn0.95 & \phs\phn1.10 &      \\ 
126.58\phn & \phs0.20\phn & \phn$-$11.9       & \phs\phn2.99 & \phs\phn5.15 & MSXE \\
           &              & \phn$-$13.1       & \phn$-$2.85  & \phn$-$6.56  &      \\ 
           &              & \phn$-$52.9       & \phs\phn1.10 & \phs\phn1.15 &      \\ 
126.83\phn & \phs0.94\phn & \phn$-$10.5       & \phs\phn2.42 & \phs\phn3.95 & MSXE \\
           &              & \phn$-$36.3       & \phs\phn1.65 & \phs\phn1.99 &      \\ 
           &              & \phn$-$54.0       & \phn$-$0.83  & \phn$-$1.03  &      \\
126.98\phn & \phs1.07\phn & \phn$-$37.0       & \phs\phn2.11 & \phs\phn3.28 & MSXE \\
           &              & \phn$-$53.9       & \phs\phn1.02 & \phs\phn1.59 &      \\ 
127.03\phn & \phs0.76\phn & \phn$-$41.9       & \phs12.02    & \phs19.70    & MSXE \\
           &              & \phn$-$51.0       & \phs\phn0.40 & \phs\phn0.77 &      \\ 
127.16\phn & $-$0.53\phn  & \phn$-$13.3       & \phn$-$2.46  & \phn$-$2.12  & MSXE \\
           &              & \phn$-$44.5       & \phs\phn5.96 & \phs12.74    &      \\ 
           &              & \phn$-$53.8       & \phs\phn1.28 & \phs\phn1.45 &      \\ 
           &              & \phn$-$56.1       & \phn$-$3.35  & \phn$-$3.49  &      \\ 
127.16\phn & \phs1.87\phn & \phn\phn$-$1.1    & \phs\phn2.11 & \phs\phn5.45 & MSXE \\
127.21\phn & \phs0.98\phn & \phn$-$43.0       & \phs\phn9.04 & \phs22.61    & MSXE \\
           &              & \phn$-$52.1       & \phs\phn2.98 & \phs\phn3.37 &      \\ 
127.45\phn & \phs0.67\phn & \phs\phn\phn9.4   & \phs\phn1.80 & \phs\phn5.93 & MSXE \\
           &              & \phn$-$52.5       & \phs\phn1.74 & \phs\phn2.69 &      \\ 
127.45\phn & \phs0.52\phn & \phn$-$34.4       & \phs\phn3.73 & \phs\phn4.60 & MSXE \\
           &              & \phn$-$50.4       & \phs\phn1.47 & \phs\phn3.13 &      \\ 
127.48\phn & \phs0.98\phn & \phn$-$39.7       & \phs\phn8.56 & \phs\phn8.99 & MSXE \\
127.49\phn & \phs1.16\phn & \phn$-$36.0       & \phs\phn0.51 & \phs\phn0.57 & MSXE \\
           &              & \phn$-$52.1       & \phs\phn0.61 & \phs\phn1.31 &      \\ 
           &              & \phn$-$58.2       & \phs\phn0.37 & \phs\phn0.80 &      \\ 
127.62\phn & \phs0.89\phn & \phn$-$41.5       & \phs\phn7.08 & \phs\phn8.04 & MSXE \\
           &              & \phn$-$51.7       & \phs\phn1.28 & \phs\phn1.61 &      \\ 
127.78\phn & \phs0.61\phn & \phs\phn\phn8.1   & \phs\phn1.22 & \phs\phn2.41 & MSXE \\
127.84\phn & \phs1.70\phn & \phn$-$55.4       & \phs\phn2.08 & \phs\phn2.40 & MSXE \\
127.86\phn & \phs1.16\phn & \phn$-$59.8       & \phs\phn2.57 & \phs\phn6.34 & MSXE \\
           &              & \phn$-$53.8       & \phs\phn1.68 & \phs\phn2.95 &      \\ 
127.98\phn & \phs4.69\phn & \phn\phn$-$8.7    & \phs\phn2.93 & \phs\phn4.41 & MSXE \\
128.17\phn &  $-$0.45\phn & \phn$-$12.7       & \phs\phn2.51 & \phs\phn4.67 & MSXE \\
           &              & \phn$-$41.7       & \phs\phn3.67 & \phs\phn5.48 &      \\ 
128.18\phn & \phs3.59\phn & \phn\phn$-$8.9    & \phs\phn2.31 & \phs\phn2.11 & MSXE \\
128.55\phn & \phs1.24\phn & \phn$-$51.1       & \phs\phn1.71 & \phs\phn3.36 & MSXE \\
           &              & \phn$-$52.9       & \phs\phn1.06 & \phs\phn0.87 &      \\ 
128.82\phn & \phs1.59\phn & \phn\phn$-$7.8    & \phs\phn1.57 & \phs\phn1.67 & MSXE \\
           &              & \phn$-$10.0       & \phs\phn1.27 & \phs\phn1.58 &      \\ 
           &              & \phn$-$55.7       & \phs\phn2.65 & \phs\phn4.02 &      \\ 
           &              & \phn$-$82.2       & \phs\phn1.71 & \phs\phn2.60 &      \\ 
129.81\phn & \phs1.43\phn & \phn$-$38.4       & \phs\phn1.86 & \phs\phn4.15 & MSXE \\
           &              & \phn$-$55.0       & \phs\phn1.21 & \phs\phn1.51 &      \\ 
           &              & \phn$-$77.0       & \phs\phn0.61 & \phs\phn0.69 &      \\ 
           &              & \phn$-$82.2       & \phs\phn0.82 & \phs\phn0.88 &      \\ 
130.00\phn & \phs1.83\phn & \phn$-$47.2       & \phs\phn3.71 & \phs\phn3.76 & MSXE \\
           &              & \phn$-$81.0       & \phs\phn1.00 & \phs\phn1.46 &      \\ 
130.13\phn & \phs1.91\phn & \phs\phn\phn2.4   & \phs\phn1.10 & \phs\phn1.86 & MSXE \\
130.18\phn & \phs0.53\phn & \phn$-$10.9       & \phs\phn2.38 & \phs\phn2.76 & MSXE \\
           &              & \phn$-$12.6       & \phs\phn1.21 & \phs\phn1.18 &      \\ 
           &              & \phn$-$45.6       & \phs\phn0.86 & \phs\phn1.10 &      \\ 
130.52\phn & \phs1.21\phn & \phn$-$66.0       & \phs\phn0.57 & \phs\phn0.83 & MSXE \\
130.61\phn & \phs2.85\phn & \phn\phn$-$7.4    & \phn$-$1.73  & \phn$-$2.33  & MSXE \\
130.82\phn &  $-$0.02\phn & \phn$-$17.1       & \phs\phn4.32 & \phs\phn5.54 & MSXE \\
           &              & \phn$-$33.0       & \phs\phn3.38 & \phs\phn4.07 &      \\ 
130.95\phn & \phs4.43\phn & \phn$-$14.1       & \phs\phn1.78 & \phs\phn2.87 & MSXE \\
130.99\phn & \phs0.56\phn & \phn$-$30.8       & \phs\phn1.91 & \phs\phn3.18 & MSXE \\
131.53\phn & \phs4.58\phn & \phn$-$33.1       & \phs\phn6.67 & \phs11.48    & MSXE \\
131.79\phn & \phs3.75\phn & \phn\phn$-$8.2    & \phn$-$1.49  & \phn$-$1.50  & MSXE \\
           &              & \phn\phn$-$8.0    & \phs\phn0.80 & \phs\phn0.38 &      \\ 
           &              & \phn$-$40.7       & \phn$-$0.66  & \phn$-$1.21  &      \\
131.81\phn & \phs1.36\phn & \phn$-$57.4       & \phs\phn1.22 & \phs\phn1.69 & MSXE \\
           &              & \phn$-$79.0       & \phs\phn3.11 & \phs\phn4.68 &      \\ 
131.94\phn & \phs1.82\phn & \phn$-$45.6       & \phs\phn1.22 & \phs\phn2.02 & MSXE \\
           &              & \phn$-$69.9       & \phs\phn0.89 & \phs\phn1.60 &      \\ 
132.12\phn & \phs1.65\phn & \phn$-$75.0       & \phs\phn0.43 & \phs\phn0.88 & MSXE \\
           &              &    $-$101.7       & \phs\phn0.37 & \phs\phn0.39 &      \\ 
132.16\phn & \phs3.84\phn & \phn$-$41.2       & \phs\phn2.30 & \phs\phn4.03 & MSXE \\
132.17\phn & \phs2.29\phn & \phn\phn$-$7.7    & \phs\phn2.69 & \phs\phn3.50 & MSXE \\
           &              & \phn$-$73.7       & \phs\phn0.67 & \phs\phn0.75 &      \\ 
132.36\phn & \phs3.98\phn & \phn$-$41.1       & \phs\phn1.80 & \phs\phn2.65 & MSXE \\
132.83\phn & \phs3.89\phn & \phn\phn$-$1.5    & \phs\phn1.24 & \phs\phn1.48 & MSXE \\
           &              & \phn\phn$-$7.9    & \phn$-$1.06  & \phn$-$1.24  &      \\ 
           &              & \phn\phn$-$9.6    & \phs\phn3.05 & \phs\phn2.71 &      \\ 
131.666    & \phs1.934    & \phn$-$41.2       & \phs\phn0.44 & \phs\phn0.44 & MSXP \\
           &              & \phn$-$79.1       & \phs\phn0.46 & \phs\phn0.72 &      \\ 
131.856    & \phs1.331    & \phn$-$56.1       & \phs\phn0.49 & \phs\phn0.72 & MSXP \\
           &              & \phn$-$78.7       & \phs\phn8.38 & \phs23.94    &      \\ 
131.826    & \phs1.364    & \phn$-$56.3       & \phs\phn0.38 & \phs\phn0.43 & MSXP \\
           &              & \phn$-$78.5       & \phs\phn3.14 & \phs\phn4.48 &      \\ 
130.294    & \phs1.654    & \phn\phn$-$3.3    & \phs\phn0.51 & \phs\phn0.66 & MSXP \\
           &              & \phn$-$54.5       & \phs\phn6.04 & \phs\phn9.95 &      \\ 
131.709    & \phs0.624    & \phn\phn$-$2.7    & \phs\phn2.21 & \phs\phn6.50 & MSXP \\
131.186    & \phs0.474    & \nodata           & \nodata      & \nodata      & MSXP \\
131.136    & \phs0.164    & \phn\phn$-$5.0    & \phs\phn2.00 & \phs\phn4.10 & MSXP \\
           &              & \phn$-$19.0       & \phs\phn0.30 & \phs\phn0.86 &      \\ 
130.416    & \phs0.266    & \phn\phn$-$9.1    & \phs\phn0.69 & \phs\phn0.44 & MSXP \\
           &              & \phn$-$11.5       & \phn$-$4.65  & $-$11.81     &      \\ 
130.896    & \phs0.419    & \phn$-$58.9       & \phs\phn2.20 & \phs\phn5.48 & MSXP \\
130.484    & \phs0.156    & \phn$-$11.5       & \phn$-$1.20  & \phn$-$1.91  & MSXP \\
           &              & \phn$-$88.7       & \phs\phn0.40 & \phs\phn0.87 &      \\ 
130.294    & \phs1.654    & \phn$-$54.5       & \phs\phn6.49 & \phs10.34    & MSXP \\
126.43\phn & \phs4.48\phn & \phs\phn\phn2.2   & \phs\phn3.82 & \phs\phn6.25 & FT96 \\
           &              & \phn$-$10.2       & \phs\phn2.23 & \phs\phn5.69 &      \\ 
127.09\phn & \phs1.86\phn & \phn\phn$-$0.4    & \phs\phn2.66 & \phs\phn3.63 & FT96 \\
           &              & \phn\phn$-$2.9    & \phs\phn1.38 & \phs\phn1.53 &      \\ 
           &              & \phn\phn$-$8.1    & \phs\phn1.00 & \phs\phn0.64 &      \\ 
127.83\phn & \phs1.67\phn & \phn\phn$-$7.3    & \phs\phn1.17 & \phs\phn1.81 & FT96 \\
           &              & \phn$-$57.7       & \phs\phn4.38 & \phs\phn4.23 &      \\ 
           &              & \phn$-$87.3       & \phn$-$1.30  & \phn$-$1.98  &      \\ 
127.91\phn & \phs0.66\phn & \phn$-$63.4       & \phs\phn6.55 & \phs19.10    & FT96 \\
128.78\phn & \phs2.01\phn & \phn$-$82.0       & \phs10.47    & \phs30.76    & FT96 \\
130.00\phn & \phs1.81\phn & \phn$-$43.5       & \phs\phn0.59 & \phs\phn0.86 & FT96 \\
           &              & \phn$-$47.2       & \phs\phn2.89 & \phs\phn2.96 &      \\ 
           &              & \phn$-$81.5       & \phs\phn0.52 & \phs\phn1.22 &      \\ 
130.29\phn & \phs1.66\phn & \phn$-$10.8       & \phs\phn0.55 & \phs\phn0.79 & FT96 \\
           &              & \phn$-$43.0       & \phs\phn0.77 & \phs\phn1.26 &      \\ 
           &              & \phn$-$54.5       & \phs\phn3.45 & \phs\phn6.05 & 
\enddata
\label{chap3-targets}
\end{deluxetable}

\end{document}